\documentstyle[12pt,epsf]{article}

\makeatletter
\newif\ifnoncomplete

\noncompletefalse

\newif\ifrefphysrev

\refphysrevfalse

\def \vol(#1,#2,#3){\ifrefphysrev{{\bf {#1}},
{#3} (19{#2})}\else{{{\bf {#1}}(19{#2}){#3}}}\fi}
\def \NP(#1,#2,#3){Nucl.\ Phys.\          \vol(#1,#2,#3)}
\def \PL(#1,#2,#3){Phys.\ Lett.\          \vol(#1,#2,#3)}
\def \PRL(#1,#2,#3){Phys.\ Rev.\ Lett.\   \vol(#1,#2,#3)}
\def \PRp(#1,#2,#3){Phys.\ Rep.\          \vol(#1,#2,#3)}
\def \PR(#1,#2,#3){Phys.\ Rev.\           \vol(#1,#2,#3)}
\def \PTP(#1,#2,#3){Prog.\ Theor.\ Phys.\ \vol(#1,#2,#3)}
\def \ibid(#1,#2,#3){{\it ibid.}\         \vol(#1,#2,#3)}

\def\thebibliography#1{
\section*{References\@mkboth
  {REFERENCES}{REFERENCES}}\list
  {[\arabic{enumi}]}{\setlength\labelwidth{2ex}
   \setlength\labelsep{0.05in} 
   \setlength\leftmargin{0.25in}
    \setlength\itemsep{0pt}
    \setlength\parsep{0pt}
   \itemsep\parskip
    \usecounter{enumi}}
    \def\newblock{\hskip .11em plus .33em minus -.22em}
    \sloppy
    \sfcode`\.=1000\relax}

\def\@bibitem#1{\item\if@filesw \immediate\write\@auxout
       {\string\bibcite{#1}{\the\c@enumi}}\fi\ignorespaces
       {\ifnoncomplete\reversemarginpar{\hspace*{-1.05in}\makebox[1in][l]
       {{\footnotesize{\sl [#1]}}}}\fi}%
       }

\def\@cite#1#2{\unskip\nobreak\relax
    {[#1]}} 

\def\citenum#1{{\def\@cite##1##2{##1}\cite{#1}}}
\def\citea#1{\@cite{#1}{}}

\newcount\@tempcntc
\def\@citex[#1]#2{\if@filesw\immediate\write\@auxout{\string\citation{#2}}\fi
  \@tempcnta\z@\@tempcntb\m@ne\def\@citea{}\@cite{\@for\@citeb:=#2\do
    {\@ifundefined
       {b@\@citeb}{\@citeo\@tempcntb\m@ne\@citea\def\@citea{,}{\bf ?}\@warning
       {Citation `\@citeb' on page \thepage \space undefined}}%
    {\setbox\z@\hbox{\global\@tempcntc0\csname b@\@citeb\endcsname\relax}%
     \ifnum\@tempcntc=\z@ \@citeo\@tempcntb\m@ne
       \@citea\def\@citea{,}\hbox{\csname b@\@citeb\endcsname}%
     \else
      \advance\@tempcntb\@ne
      \ifnum\@tempcntb=\@tempcntc
      \else\advance\@tempcntb\m@ne\@citeo
      \@tempcnta\@tempcntc\@tempcntb\@tempcntc\fi\fi}}\@citeo}{#1}}
\def\@citeo{\ifnum\@tempcnta>\@tempcntb\else\@citea\def\@citea{,}%
  \ifnum\@tempcnta=\@tempcntb\the\@tempcnta\else
   {\advance\@tempcnta\@ne\ifnum\@tempcnta=\@tempcntb \else \def\@citea{--}\fi
    \advance\@tempcnta\m@ne\the\@tempcnta\@citea\the\@tempcntb}\fi\fi}

\def\affiliation#1{\cr
\makebox[0in]{\parbox{8in}{\begin{center} {\sl #1}\end{center}}} \cr}
\def\@affiliation{}

\def\and{\cr \makebox[0in]{\rule[-1cm]{0mm}{1cm}and } \cr}

\def\maketitle{\par
 \begingroup
 \def\thefootnote{\fnsymbol{footnote}}
 \def\@makefnmark{\hbox
 to 0pt{$^{\@thefnmark}$\hss}}
 \if@twocolumn
 \twocolumn[\@maketitle]
 \else \newpage
 \global\@topnum\z@ \@maketitle \fi\thispagestyle{plain}\@thanks
 \endgroup
 \setcounter{footnote}{0}
 \let\maketitle\relax
 \let\@maketitle\relax
 \gdef\@thanks{}\gdef\@author{}\gdef\@title{}
 \gdef\@affiliation{} \let\affiliation\relax	%
 \let\thanks\relax}

\def\@maketitle{\newpage
 \null
 \vskip 0em plus 2em minus 0em     
 \ifx\@date\@empty\else
   \begin{flushright}
    {\ifnoncomplete(\today)
     \else{{\normalsize \@date}\\}\fi}      
   \end{flushright}
   \vskip 3em plus 2em minus 2em   
 \fi
 \begin{center}
  {\frtnsfb \@title \par}     
  \vskip 3em plus 1em minus 1.5em  
  {
   \lineskip .5em plus 0em minus .3em   
   \begin{tabular}[t]{c}\@author
   \end{tabular}\par}
\end{center}
 \par
 \vskip 6em plus 2em minus 4em}     

\def\abstract{\if@twocolumn
\section*{Abstract}
\else \normalsize
\fi}

\def\endabstract{\if@twocolumn\fi\par\clearpage}

\def\section{\@startsection {section}{1}{\z@}{3.5ex plus 1ex minus
    .2ex}{2.3ex plus .2ex}{\normalsize\bf}}
\def\subsection#1{\subsectioncom{\sc{#1}}}
\def\subsectioncom{\@startsection{subsection}{2}{\z@}
    {3.25ex plus 1ex minus .2ex}{1.5ex plus .2ex}{\small}}
\def\subsubsection{\@startsection{subsubsection}{3}{\z@}{3.25ex plus
1ex minus .2ex}{1.5ex plus .2ex}{\small}}

\def\@addmarginpar{\@next\@marbox\@currlist{\@cons\@freelist\@marbox
    \@cons\@freelist\@currbox}\@latexbug\@tempcnta\@ne
    \if@twocolumn
        \if@firstcolumn \@tempcnta\m@ne \fi
    \else
      \if@mparswitch
         \ifodd\c@page \else\@tempcnta\m@ne \fi
      \fi
      \if@reversemargin \@tempcnta -\@tempcnta \fi
    \fi
    \ifnum\@tempcnta <\z@  \global\setbox\@marbox\box\@currbox \fi
    \@tempdima\@mparbottom \advance\@tempdima -\@pageht
       \advance\@tempdima\ht\@marbox \ifdim\@tempdima >\z@
      \else\@tempdima\z@ \fi
    \global\@mparbottom\@pageht \global\advance\@mparbottom\@tempdima
       \global\advance\@mparbottom\dp\@marbox
       \global\advance\@mparbottom\marginparpush
    \advance\@tempdima -\ht\@marbox
    \global\ht\@marbox\z@ \global\dp\@marbox\z@
    \vskip -\@pagedp \vskip\@tempdima\nointerlineskip
    \hbox to\columnwidth
      {\ifnum \@tempcnta >\z@
          \hskip\columnwidth \hskip\marginparsep
        \else \hskip -\marginparsep \hskip -\marginparwidth \fi
       \box\@marbox \hss}
    \vskip -\@tempdima
    \nointerlineskip
    \hbox{\vrule \@height\z@ \@width\z@ \@depth\@pagedp}}

\def\ref#1{
    \@ifundefined{r@#1}{{#1}\@warning{Reference `#1'
    on page \thepage \space
    undefined}}{\edef\@tempa{\@nameuse{r@#1}}\expandafter
    \@car\@tempa \@nil\null}}
\def\refn#1{\@ifundefined{r@#1}{{#1}\@warning{Reference `#1'
    on page \thepage \space
    undefined}}{\edef\@tempa{\@nameuse{r@#1}}\expandafter
    \@car\@tempa \@nil\null}}

\def\endequationl{\eqno \@eqnnum 
$$\global\@ignoretrue}

\def\eqnarray{\stepcounter{equation}\let\@currentlabel=\theequation
\global\@eqnswtrue
\global\@eqcnt\z@\tabskip\@centering\let\\=\@eqncr
$$\arraycolsep\z@
\halign to \displaywidth\bgroup\@eqnsel\hskip\@centering
  $\displaystyle\tabskip\z@{##}$&\global\@eqcnt\@ne
  \hskip 2\arraycolsep \hfil$\displaystyle{{}##{}}$\hfil
  &\global\@eqcnt\tw@ \hskip 2\arraycolsep
  $\displaystyle\tabskip\z@{##}$\hfil
   \tabskip\@centering&\llap{##}\tabskip\z@\cr}
\def\mmodetrue{\mmode=\iftrue}

\def\eqnarrayl#1{\stepcounter{equation}\let\@currentlabel=\theequation
\label {#1}
\global\@eqnswtrue
\global\@eqcnt\z@\tabskip\@centering\let\\=\@eqncr
$$\arraycolsep\z@
\halign to \displaywidth\bgroup\@eqnsel\hskip\@centering
  $\displaystyle\tabskip\z@{##}$&\global\@eqcnt\@ne
  \hskip 2\arraycolsep \hfil$\displaystyle{{}##{}}$\hfil
  &\global\@eqcnt\tw@ \hskip 2\arraycolsep
  $\displaystyle\tabskip\z@{##}$\hfil
   \tabskip\@centering&\llap{##}\tabskip\z@\cr}
\def\label#1{
\@bsphack\if@filesw {
{\ifnoncomplete{\makebox[1in][r]{\footnotesize{\sl [#1]}}}\fi}%
\let\thepage\relax
   \xdef\@gtempa{\write\@auxout{\string
      \newlabel{#1}{{\@currentlabel}{\thepage}}}}
}\@gtempa
   \if@nobreak \ifvmode\nobreak\fi\fi\fi\@esphack}

\def\newlabel#1#2{
\@ifundefined{r@#1}{}{\@warning{Label `#1' multiply
   defined}}\global\@namedef{r@#1}{#2}}

\def\endeqnarrayl{\@@eqncr\egroup
      \global\advance\c@equation\m@ne$$\global\@ignoretrue}

\newif\if@numbersec \@numbersectrue
\def\appendix{\par\clearpage
  \setcounter{section}{0}
  \setcounter{subsection}{0}
  \def\thesection{\Alph{section}}
  \def\thesubsection{\arabic{subsection}}
  \@ifstar{\def\@sectname{Appendix}\@numbersecfalse}
          {\def\@sectname{Appendix~}\@numbersectrue}}

\def\thefigures#1{\par\clearpage\section*{Figures\@mkboth
  {FIGURES}{FIGURES}}\list
  {Fig.~\arabic{enumi}.}{\labelwidth\parindent\advance\labelwidth -\labelsep
      \leftmargin\parindent\usecounter{enumi}}}

\def\thetables#1{\par\clearpage\section*{Tables\@mkboth
  {TABLES}{TABLES}}\list
  {Table~\arabic{enumi}.}{\labelwidth-\labelsep
      \leftmargin0pt\usecounter{enumi}}}

\def\@sect#1#2#3#4#5#6[#7]#8{\ifnum #2>\c@secnumdepth
     \def\@svsec{}\else
     \refstepcounter{#1}\edef\@svsec{\ifnum #2=1 \@sectname
         \if@numbersec\csname the#1\endcsname\fi.\else
         \csname the#1\endcsname.\fi
        \hskip 1em }\fi
     \@tempskipa #5\relax
      \ifdim \@tempskipa>\z@
        \begingroup #6\relax
          \@hangfrom{\hskip #3\relax\@svsec}{\interlinepenalty \@M #8\par}
        \endgroup
       \csname #1mark\endcsname{#7}\addcontentsline
         {toc}{#1}{\ifnum #2>\c@secnumdepth \else
                      \protect\numberline{\csname the#1\endcsname}\fi
                    #7}\else
        \def\@svsechd{#6\hskip #3\@svsec #8\csname #1mark\endcsname
                      {#7}\addcontentsline
                           {toc}{#1}{\ifnum #2>\c@secnumdepth \else
                             \protect\numberline{\csname the#1\endcsname}\fi
                       #7}}\fi
     \@xsect{#5}}

\def\@sectname{}

 \def\thefootnote{\fnsymbol{footnote}}

\def \@magscale#1{ scaled \magstep #1}
\font\frtnsfb = cmssbx10 \@magscale2 
\newcount\ieq
\newcount\jeq
\newcount\keq
\def \eq{
\multiply\ieq by 2
\jeq=\ieq
\divide\jeq by 4
\multiply\jeq by 4
\ifnum\ieq=\jeq \end{eqnarray} \keq=1 
\else
\keq=2 \begin{eqnarray} \fi
\ieq=\keq
}

\def \mathbox(#1){\invisible\ifmmode{{#1}}\else{\mbox{${#1}$}}\fi}
\def \mbf(#1){\mbox{\boldmath{$#1$}}}
\def \abs(#1){\mathbox(\left|{#1}\right|)}
\def \bracket(#1){\mathbox(\left\langle{#1}\right\rangle)}
\def \brav(#1){\mathbox(\langle {#1}|)}
\def \cg(#1,#2,#3,#4,#5,#6){\mathbox({(#1\,#2\,#3\,#4|#5\,#6)})}
\def \comm(#1,#2){\mathbox(\left[{#1},{#2}\right])}
\def \dfdx(#1,#2){\mathbox(\frac{{\rm d}{#1}}{{\rm d}{#2}})}
\def \delfdelx(#1,#2){\mathbox(\frac{\partial{#1}}{\partial{#2}})}
\def \inprod(#1,#2){\mathbox({(#1\cdot #2)})}
\def \inprodij(#1){\mathbox({\inprod(#1_i,#1_j)})}
\def \intd(#1,#2){\mathbox({\int^#1_#2 \; \rmd})}
\def \eps(#1){\mathbox(\epsilon_{#1})}
\def \half(#1){\mathbox(\frac{#1}{2})}

\def \ketv(#1){\mathbox(|{#1}\rangle)}
\def \matele(#1,#2,#3){\mathbox(\left\langle {#1}|\,{#2}\,|{#3}\right\rangle)}
\def \mateled(#1,#2,#3){\mathbox(\left\langle
{#1}||\,{#2}\,||{#3}\right\rangle)}
\def \hatmbf(#1){\mathbox({\hat{\mbf({#1})}})}
\def \ninej(#1,#2,#3,#4,#5,#6,#7,#8,#9){\mathbox(\left\{\matrix
     {#1&#2&#3\cr#4&#5&#6\cr#7&#8&#9\cr}\right\})}
\def \rtov(#1,#2){\mathbox(\sqrt{{#1\over #2}})}
\def \sixj(#1,#2,#3,#4,#5,#6){\mathbox(\left\{\matrix
     {#1&#2&#3\cr#4&#5&#6\cr}\right\})}
\def \third(#1){\mathbox(\frac{#1}{3})}
\def \Trace(#1){\mathbox({\hbox{Tr} \left\{#1\right\}})}
\def \outprod(#1,#2){\mathbox({(#1\times #2)})}

\def \calA{\mathbox({\cal A})}

\def \etal{{\it et al.}}
\def \etc{{\it etc.}}

\def \ie{{\it i.e.}}
\def \invisible{\mbox{$\rule{0mm}{1mm}$}}

\def \Lam{\mbox{$\Lambda$}}

\def \rmd{{\rm d}}

\def \Sig{\mbox{$\Sigma$}}

\makeatother

\def \abs #1{\mbox{$\left|{#1}\right|$}}
\def \bracket<#1>{\mbox{$\langle {#1}\rangle$}}
\def \brav #1|{\mbox{$\langle {#1}|$}}
\def \cg(#1,#2,#3,#4,#5,#6){\mbox{$(#1,#2,#3,#4|#5,#6)$}}
\def \comm[#1,#2]{\mbox{$\left[{#1},{#2}\right]$}}
\def \ddt #1{\ifmmode{{\partial #1\over\partial t}} \else{${\partial
#1\over\partial t}$}\fi}

\def \dotp(#1.#2){\mbox{$(#1\cdot #2)$}}

\def \ketv #1>{\mbox{$|{#1}\rangle$}}
\def \mate<#1|#2|#3>{\mbox{$\langle {#1}|\,{#2}\,|{#3}\rangle$}}
\def \mated<#1||#2||#3>{\mbox{$\langle {#1}||\,{#2}\,||{#3}\rangle$}}
\def \mvec #1{\mbox{\boldmath{${#1}$}}}
\def \rtov(#1/#2){{\ifmmode{{\sqrt{\frac{#1}{#2}}}}
\else{{$\sqrt{\frac{#1}{#2}}$}}\fi}}
\def \sixj(#1,#2,#3,#4,#5,#6){\mbox{$\left\{\matrix
{#1&#2&#3\cr#4&#5&#6\cr}\right\}$}}
\def \Trace[#1]{\mbox{${\hbox{Tr} \left\{#1\right\}}$}}
\def \xp(#1.#2){\mbox{${(#1\times #2)}$}}

\def \calA{\mbox{${\cal A}$}}

\def \etal{{\it et al.}}
\def \etc{{\it etc.}}
\def \half{\mbox{$\frac{\rm 1}{\rm 2}$}}

\def \ie{{\it i.e.}}
\def \III{{\rm I\thinspace I\thinspace I}}

\def \KYazaki{K.\ Yazaki}

\def \Lam{\mbox{$\Lambda$}}

\def \MOka{M.\ Oka}

\def \rhat{\ifmmode{\hat{\mvec r}}\else{$\hat{\mvec r}$}\fi}
\def \rmd{{\rm d}}

\def \Sachiko{S.\ Takeuchi}
\def \sigsig{\mbox{$(\sigv_i\cdot\sigv_j)$}}
\def \sigv{\mbox{$\vec\sigma$}}
\def \Sig{\mbox{$\Sigma$}}

\def \third{\mbox{$\frac{1}{3}$}}

\def \rmd{{\mbox{{\rm d}}}}

\def \ddt #1{\ifmmode{{\partial #1\over\partial t}} \else{${\partial
#1\over\partial t}$}\fi}
\def \doublet(#1,#2){{\mbox{$\left( {\matrix{#1\cr#2\cr}} \right) $}}}
\def \triplet(#1,#2,#3){{\left(\matrix{#1\cr#2\cr#3\cr}\right)}}
\def \sixj(#1,#2,#3,#4,#5,#6)
{{\mbox{$\left\{\matrix{#1&#2&#3\cr#4&#5&#6\cr}\right\}$}}}

\def \ninej(#1,#2,#3,#4,#5,#6,#7,#8,#9)
    {\mbox{$\left\{\matrix {#1&#2&#3\cr#4&#5&#6\cr#7&#8&#9\cr}\right\}$}}

\def\wave{\simeq}
\def\Lam{\Lambda}
\def\Sig{\Sigma}

\begin{document}

\rightline{TIT/HEP-305/NP}
\rightline{October, 1995}

\begin{center}
{\Large\bf QCD Analysis of the H dibaryon%
\footnote{talk presented at the International Symposium on Exotic
Atoms and Nuclei, June 7--10, 1995, Hakone, Japan}}\\
{\large Makoto Oka}\\
{\sl Department of Physics, Tokyo Institute of Technology}\\
{\setlength{\baselineskip}{12pt}
{\sl Meguro, Tokyo 152 Japan}\\
{\sl e-mail: oka@th.phys.titech.ac.jp}\\
}
\end{center}
\centerline{\bf Abstract}
\noindent The status of theoretical studies of the $H$ dibaryon is
reviewed.  Some recent developments including the effect of the
instanton induced interaction and the QCD sum rule results are
discussed in detail.

\section{Introduction}

\begin{figure}[thb]
\centerline{\epsfbox{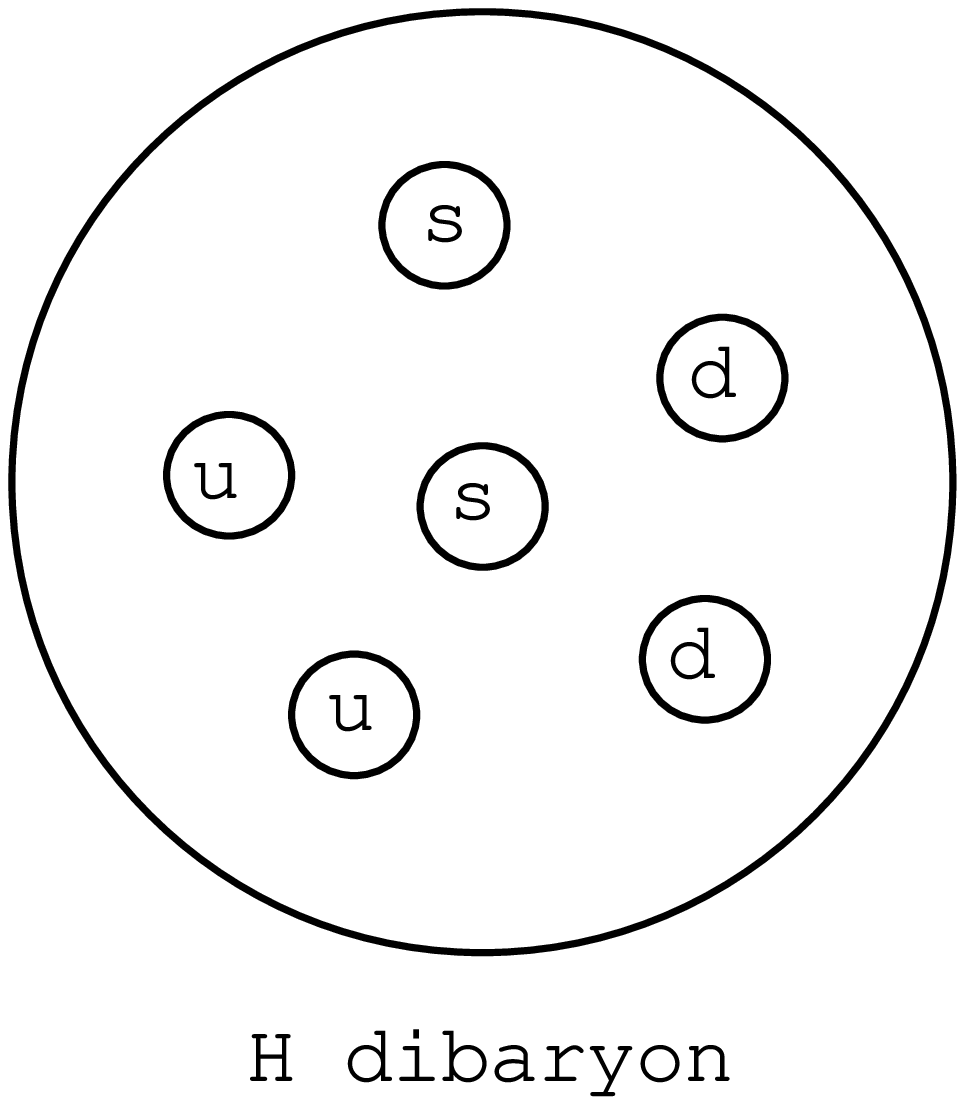}}
\caption{$H$ dibaryon with $J^{\pi}= 0^+$, $I=0$, $B=2$,
$S=-2$ and $Y=0$. }
\end{figure}

$H$ is a $J=0$ dibaryon predicted in the flavor $SU(3)_f$ singlet
representation (figure 1).
It has strangeness $-2$, isospin 0 and hypercharge $Y=B+S=0$.
In the valence quark model it consists of two $u$, two $d$ and two $s$
quarks.  $H$ is stable against strong decays if its mass is below
the two baryon thresholds, $\Lambda\Lambda$ (2231MeV) --
$N\Xi$ ($n\Xi^0$: 2254MeV, $p\Xi^-$: 2260MeV) -- $\Sigma\Sigma$
($\Sigma^0\Sigma^0$: 2385 MeV, $\Sigma^+\Sigma^-$: 2387 MeV).  The binding
energy of $H$ is measured from the lowest two baryon threshold,
$\Lambda\Lambda$.
As a six-quark object, this is truly an EXOTIC hadron, whose existence
alone is of great significance in hadron physics.

Experimental searches of $H$ have continued for some time
and yet no evidence of deeply bound states is found\cite{AGS}.  On the
contrary,
recent $(K^-,K^+)$ reaction experiments on emulsion targets identified
a few candidates of double-hypernuclei with
binding energy of less than 20 MeV{\cite{Imai}}.
Because such a double hypernucleus would make a strong-interaction
transition into $H$ and an ordinary nucleus, its existence kills the
possibility of deeply bound $H$.

In this report, I am going to review the status of theoretical
analysis of $H$.
After a brief review of the history, some recent progresses
are discussed in detail.


\section{Brief History}
In 1977,  Jaffe predicted the $H$ dibaryon in the MIT bag
model{\cite{Jaffe}} and since
then almost all possible models of hadron dynamics have been applied
to this problem.
Yet the conclusion is not reached yet.  The mechanism for
the $H$ binding in the bag model is simple.  It is due to a strong
attraction of color-magnetic gluon exchanges between quarks.
The $n$-quark state in the flavor $SU(3)$ representation
$[f]$ acquires the color-magnetic gluon interaction energy given by
\begin{equation}
	E_{cm}(S,[f]) = E_0 \,\left[n(n-10) + C_2[f] +{4\over 3}
		\,S(S+1) \right]
\end{equation}
where $S$ is the total spin and $C_2{[f]}$ denotes the value of the quadratic
Casimir operator, such as
$C_2=0$ for the singlet, 12 for the octet and 24 for the decuplet.
The overall constant $E_0$ is given by a spatial integration of the
quark wave functions.
Instead we can estimate $E_0$ from the $N-\Delta$ mass
splitting, assuming that the same interaction gives the full
splitting and that the bag radius is independent of $n$.
(The latter is not valid for the bag model, while it is justified for
the potential model of confinement.  The mechanism of confinement
gives an ambiguity here.)
As the $N-\Delta$ energy difference due to the color-magnetic
gluon is  $16\times E_0$, we find $E_0= (M_{\Delta}-M_N)/16
\wave 18$ MeV.
The $E_{cm}$ gives the minimum, $E_{cm} = -24 E_0$,  at $S=0$ and
$C_2=0$, that corresponds to the
$H$ dibaryon (flavor singlet, spin zero).
Thus our estimation $E_{cm}(H) \wave - 450$ MeV shows that the
color-magnetic gluon exchange interaction strongly favors the $H$
dibaryon.

Soon after the first prediction of $H$ was made,
it was noticed that two other effects are important
in determining the binding energy of the $H$ dibaryon.
One is the effect of confinement.
The difference of the bag volume energy
between the hyperon $\Lambda$ and $H$.  This, in fact, causes the most
serious ambiguity in predicting the $H$ dibaryon mass.   In the bag
model, we have never tested the bag volume energy term in the system with
more than three quarks.  Because the bag (surface) is not treated
dynamically, one cannot describe the hadron-hadron interaction in the bag
model properly.

The other important effect is the Pauli exclusion principle among the valence
quarks.  For the nucleon-nucleon interaction, one finds that the Pauli
exclusion gives repulsion, while introduction of the strange quarks
reduces the ``exclusiveness'' and thus the repulsion is expected to be
reduced in the $\Lambda-\Lambda$  system.

In order to study the effect of confinement and the Pauli exclusion
principle, the quark cluster model{\cite{QCM}} based on the
nonrelativistic quark model with potential confinement
was applied to the two baryon systems{\cite{OSY}}.
This approach has a strong advantage that the coupling of $H$ to the
two-baryon systems, $\Lambda-\Lambda$, $N\Xi$ and $\Sigma\Sigma$, can
be taken into account consistently with the two baryon dynamics.
The quark cluster model is successful in describing the short distance
part of the nuclear force and thus application to the two hyperon
systems is a straightforward extension.

The six-quark structure of $H$ is represented by a resonating group method
wave function as
\begin{equation}
	\Psi_H(1-6) = \sum_{(BB')=(\Lam\Lam),(N\Xi),(\Sig\Sig)}
	 \calA \,\left[\phi_B(123)\,\phi_{B'}(456)\,\chi_{BB'}\right]
\protect\label{eq:RGMwf}
\end{equation}
where $\phi_B$'s are the internal quark wave functions of the baryon
and $\chi_{BB'}$ describes the relative motion of the $(BB')$ channel.
$\calA$ is the antisymmetrization operator for the six quarks.
The resonating group method is employed to solve the Schr\"odinger
equation. The full antisymmetrization is taken into account and
induces the quark exchange interaction between the baryons.

\begin{figure}[thb]
\centerline{\epsfbox{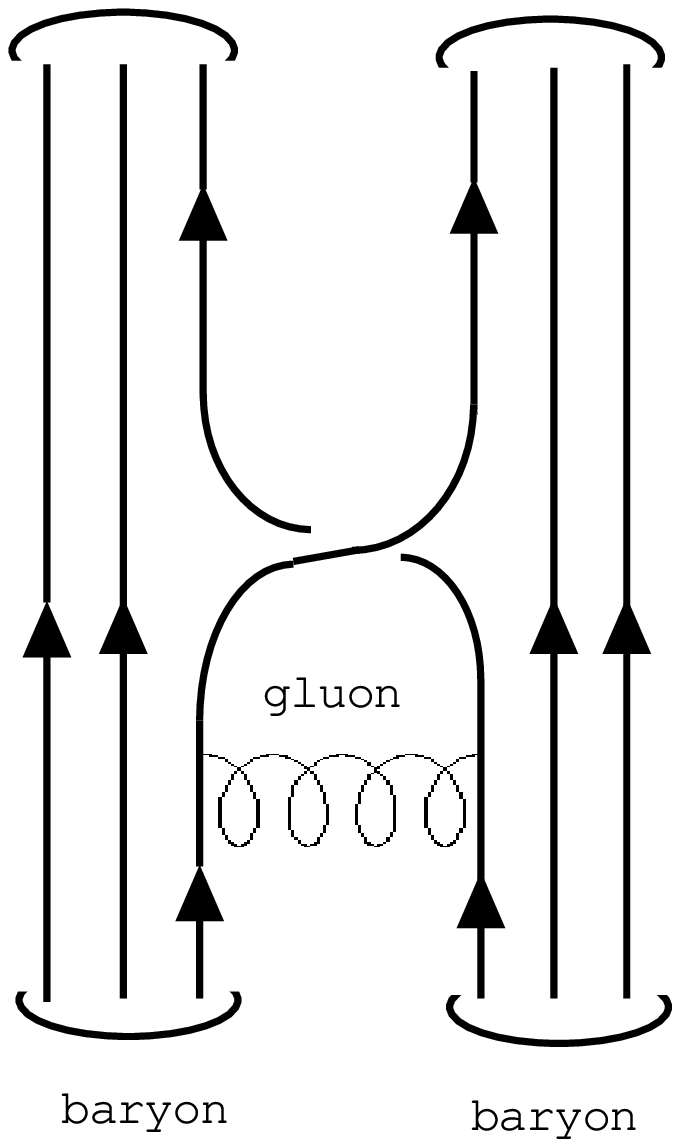}}
\caption{A quark exchange diagram associated with a gluon exchange.}
\end{figure}

In 1983,  Oka-Shimizu-Yazaki{\cite{OSY}} found that the quark exchange
diagrams (figure 2) associated with the color-magnetic gluon exchange yield a
strong attraction to the flavor-singlet two-baryon state.
\begin{equation}
   \ketv H> \simeq {1\over \sqrt{8}}\, \left[\calA \ketv \Lam\Lam>
	+\sqrt{4} \calA\ketv N\Xi> - \sqrt{3} \calA\ketv\Sig\Sig>
	\right]
\protect\label{eq:singlet}
\end{equation}
We, however, found that the quark exchange
interaction alone  cannot make $H$ bound.
It is, however, found that $H$ couples most strongly to
$N\Xi$ and that a $N\Xi$ bound state will appear in the
$\Lambda\Lambda$ continuum spectrum as a sharp resonance state.
The wave function at the resonance shows the flavor singlet structure,
eq.(\ref{eq:singlet}).
Further studies in the quark cluster model have suggested that the
long-range meson
exchange interactions may give enough attraction to make a bound $H$,
while it was pointed out that the choice of the confinement is crucial in
predicting the binding energy{\cite{Tub,koike}}.

\medskip
The Skyrme model of baryons, which is based on the chiral symmetry in the
mesonic effective lagrangian, was also applied to the $H$
problem{\cite{JKorpa}}.  A new
type of the topological soliton configuration was proposed to describe
a compact $B=2$  state, which is called the $SO(3)$
Skyrmion{\cite{Bala}}.  This
configuration has the right properties for the $H$ dibaryon and predicts a
deeply bound $H$ state.   It is however, not well understood how quantum
corrections are treated.  Especially, couplings of two baryon
states seem important but it is not included in the $H$ mass calculation.

\medskip
The lattice QCD is the most promising approach to the exact solution of
QCD at low energies.  The time-like correlator of $H$  on the lattice
was calculated but several inconsistent results were presented so
far{\cite{Iwasaki}}.
We suspect that the lattice size is not large enough to contain
the whole $H$, especially when the binding energy is small and two
baryon states couple to $H$ strongly.

\medskip
Besides the existence and the binding energy of $H$, the most important
question regarding $H$ is how compact it is.
Once $H$ is identified, then we must determine whether $H$ is like a
compact 6-quark object or just like a $\Lam\Lam$ bound state.  Theory
predictions are again quite diverse from a compact object to a loosely
bound two baryons.

\section{Effects of Axial U(1) Anomaly}

The prediction of the $H$ mass relies on the validity of the
quark model description of hadrons (and hadronic interactions).
A simple hamiltonian with a quark confinement and a one-gluon exchange
interaction made a great success in the meson and baryon spectroscopy.
A few exceptional cases include the lowest pseudoscalar mesons, such as
$\pi$, $K$, $\eta$ and $\eta'$ mesons.
The octet mesons, $\pi$, $K$ and $\eta$ are generally considered as the
Nambu-Goldstone (NG) bosons of the spontaneous  breaking of the chiral
$SU(3)\times SU(3)$ symmetry.
The nonrelativistic quark model description of these NG mesons are
reasonably good,
maybe except for the pion, which is so deeply bound that the
one-gluon exchange force may not explain the full binding energy.

Weinberg pointed out that the ninth member of the pseudoscalar nonet, $\eta'$,
is too heavy to be  regarded as a NG boson{\cite{Weinberg}}.
Indeed, the axial $U(1)_A$ symmetry is explicitly broken in the
quantum theory (due to an anomaly) and thus the ninth NG boson  does not exist.
Therefore the large mass of $\eta'$ should also be
explained by an interaction that causes the  $U(1)_A$ breaking.

In the quark model description, the effect of the $U(1)_A$ breaking can be
represented by the so-called  instanton induced interaction (\III),
which is
derived by 't Hooft considering a coupling of light quarks to
instanton field configurations{\cite{KM,tH}}.
The quark-instanton coupling induces an effective 6-quark
vertex shown in the figure 3(a).  This interaction changes the chirality of
each light flavor from $L$ to $R$ and is antisymmetric in the flavor
indices.  (This is the reason why this interaction is often called the
determinant interaction.)
The strength of \III\ can be determined by the $\eta-\eta'$ mass
difference, which comes partly from the flavor $SU(3)$ breaking and
mixing and partly from the effect of the $U_A(1)$ breaking term.
We find that the two effects are of the same order{\cite{SO}}.
Recent analysis of the $\eta \to \gamma\gamma$ decay also suggests a
significant strength of \III{\cite{TakizawaO}}.

\begin{figure}[thb]
\centerline{\epsfbox{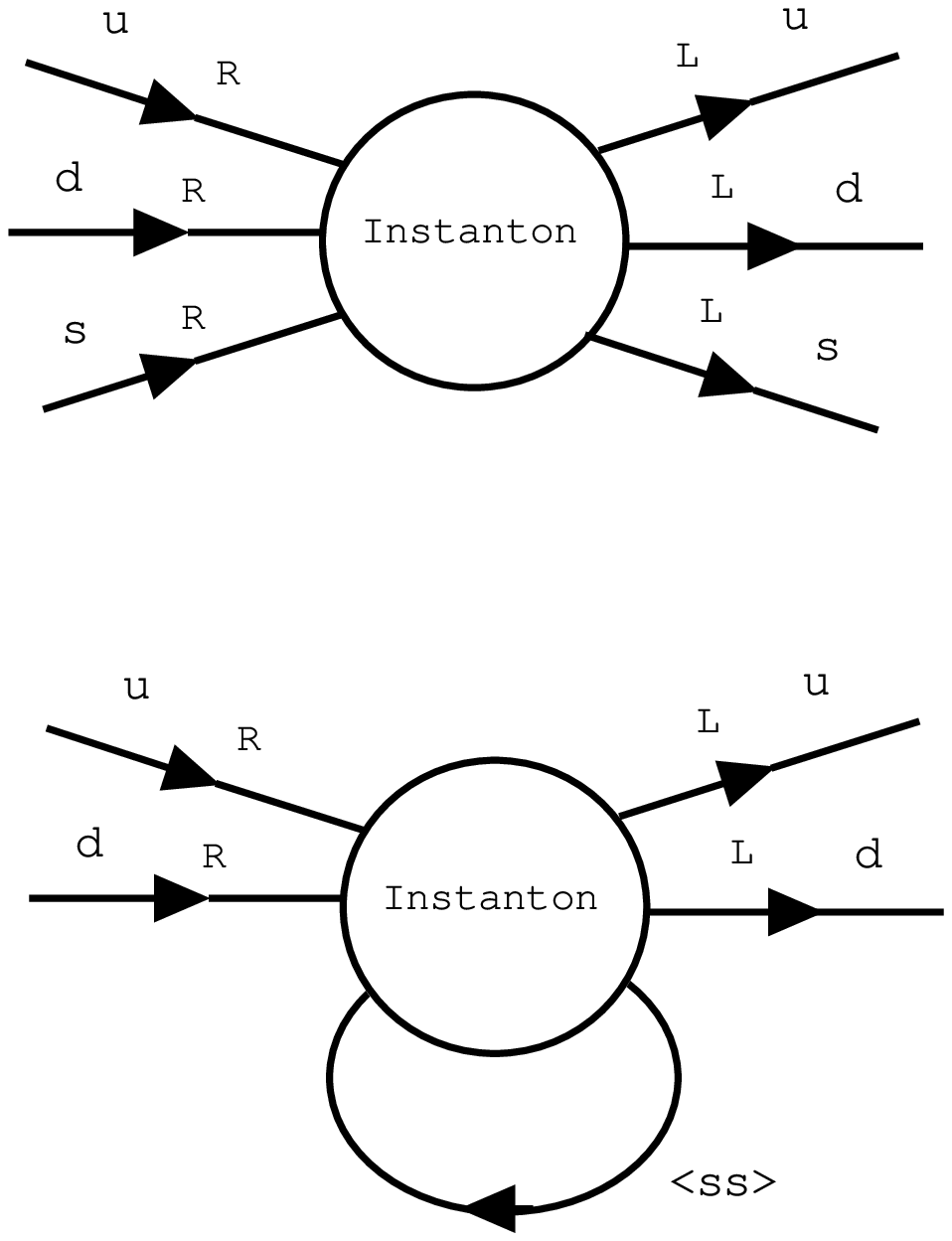}}
\caption{(a) Instanton induced interaction for the flavor
antisymmetric $u-d-s$ system.  (b) Two-body \III\ for the $I=0$ $u-d$
system.}
\end{figure}

In applying this interaction to the baryon problem, we employ the
nonrelativistic valence quark model, and
reduce \III\ into a nonrelativistic form{\cite{SO}}.
We obtain a two-body
\III\ (figure 3(b)) given by
\begin{equation}
	V^{(2)}_{\III} = V_0^{(2)} \sum_{i<j} {15\over 16} {\calA}^{(2)}_f
	\,\delta(\vec r_{ij})
	\left[1-{1\over 5}\, \sigsig \right]
\protect\label{eq:III2}
\end{equation}
and a three-body interaction (figure 3(a)),
\begin{equation}
	V^{(3)}_{\III} = V_0^{(3)} \sum_{(i,j,k)} {189\over 160}{\calA^{(3)}}_f
	\,\delta(\vec r_{ij})\delta(\vec r_{jk})
	\left[1-{1\over 7}\, \{\sigsig + {\rm permutations}\}
	\right]
\protect\label{eq:III3}
\end{equation}
where ${\cal A}^{(n)}_f$ stands for the antisymmetrization of $n$
quarks in the flavor space.
Thus this interaction is nonzero only for the flavor antisymmetric
combination of quarks.
The strength of the two-body \III\ is determined by that of the three body
\III\ by
$V_0^{(2)} = V_0^{(3)} \bracket<\bar{q} q>/2$.

It is easy to see that the three-body instanton induced interaction
(\ref{eq:III3}) vanishes in the ground state baryons because they are
not in the flavor singlet representation.
The two-body interaction (\ref{eq:III2}) gives a contribution for the
flavor antisymmetric pairs of quarks.
We find that the spin structure of the two-body \III\ is identical to
the color-magnetic gluon exchange and therefore it also explains the
hyperfine splittings of the baryon spectrum, such as the $N-\Delta$
splitting{\cite{Rosner-S,OS}}.  The \III\ strength determined by the
$\eta-\eta'$
spectrum, indeed, explains 30-40\% of the hyperfine splitting.

The three-body \III\ plays a significant role in the $H$ system.
$H$ contains two sets of antisymmetric $u-d-s$ quarks.
We performed the quark cluster model analysis of $H$ including the
\III\ term in the quark hamiltonian.
We find that the
contribution of the three-body \III\ to the $H$ is strongly repulsive,
while the two-body \III\ gives moderate attraction.
The net result amounts to
pushing the $H$ mass up by about $40-50$ MeV.
It is easy to understand that the three-body \III\ is repulsive.
The ratio of the two-body \III\ and three-body \III\ is proportional to the
quark condensate in the vacuum that is negative.  The two-body \III\
is attractive and induces the quark condensate.

We performed the quark cluster model calculation for $H${\cite{SO}} including
the quark exchange interaction, the effective meson exchange
interaction  and the instanton induced interaction.
We found that the attraction due to the two-body \III\ is mostly
absorbed into the meson exchange interaction when it is adjusted to
the $NN$ scattering data.
Thus the effect of the three-body \III\ gives a net repulsion.
Our final conclusion is that $H$ is
either barely bound or unbound, depending on how strong the \III\ is.
Even if it is not bound, it is still
possible to have  a $\Lam\Lam$ resonance state below the $N\Xi$ threshold.

\section{QCD Sum Rule analysis}

The QCD sum rule is a novel way of studying hadron spectrum and
properties directly from QCD{\cite{SVZ}}.  The sum rule takes
advantage of  analyticity
of current correlators and relates the asymptotic free region of QCD
to the nonperturbative physical region.  On one side (theoretical
side) the correlator is calculated perturbatively for a large
Euclidean momentum carried by the current and then the result is
analytically continued to the physical spectrum region.  It requires
some matrices of quark-gluon local operators in the vacuum, which are
determined either by other sum rules or by imposing consistency of
the sum rule.  On the phenomenological side, the physical spectral
density is parameterized in a form with discrete poles and continuum
parts, whose parameters (position of the poles, coupling strengths,
thresholds for the continuum, \etc) are determined so as to coincide
with the theoretical side extrapolated analytically from the deep
Euclidean region.
The process of the analytical
continuation may often be subtle because the physical parameters are
not so sensitive to the short distance behavior of the correlator.
The Borel sum and the finite-energy sum rule are two popular
techniques employed so that one can find the most appropriate weight
function in comparing the phenomenological and theoretical sides.

\begin{figure}[thb]
\centerline{\epsfbox{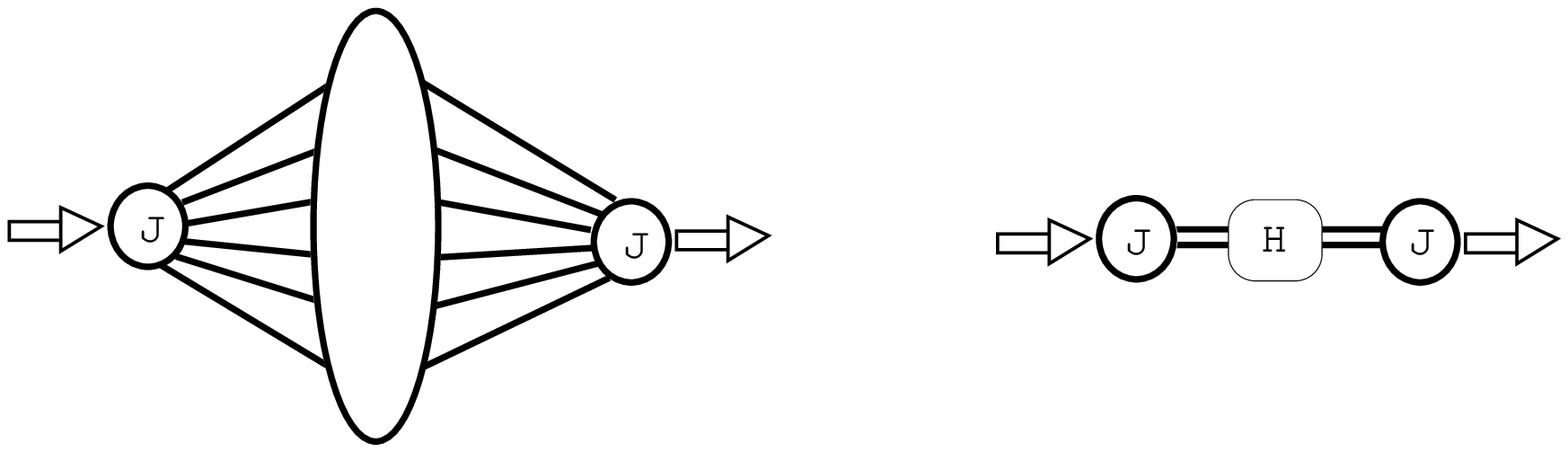}}
\caption{QCD sum rule for the $H$ current correlator.}
\end{figure}

Recently we have applied the QCD sum rule to the $H$ dibaryon
problem.
We construct the interpolating current for the $H$ dibaryon as a
product of two currents representing the flavor-octet baryons,
\begin{equation}
    J_H(x) = \sum_{\alpha=1-8} J_B^{\alpha} C\gamma_5 J_B^{\alpha}
\end{equation}
where $C$ is the charge conjugation operator and $\alpha$ labels the
flavor octet members.
We apply the QCD sum rule for the current correlator (figure 4) defined by
\begin{equation}
    \Pi_H(x) = \bracket<T[ J_H (x) J_H^{\dagger}(0)]>
\end{equation}
Details of this calculation should be referred to ref.{\cite{HSR}}


It happens that the sum rule cannot effectively fix the continuum
threshold, and thus the prediction has a large ambiguity.
We therefore compare the $H$ sum rule with a similar sum rule for the
``dinucleon'' $D$, which is a hypothetical bound state of two
nucleons (protons) in the spin singlet state.
Comparison of those two states is easy because they have the same spin
and thus similar
current structures.
To our surprise, we have found that those two sum rules are almost
identical in the $SU(3)$ limit and therefore predict nearly the same
masses for $H$ as the $^1S_0$ di-nucleon $D$, which is unbound experimentally
albeit close to be bound.
This degeneracy of $H$ and $D$ has its origin in the similarity
of the theoretical sides in the $SU(3)$ limit.

Effects of $SU(3)$ breaking are taken into account as terms
proportional to the strange quark mass and the difference of
$\bracket<\bar s s>$ and $\bracket<\bar u u>$.
Then we find that the mass of $H$ is about ($2.19\pm 0.07$) GeV, the
central value of which is about 40 MeV below the $\Lam\Lam$ threshold.
Because the result is extremely sensitive to the choice of the
continuum threshold energy, the number should not be taken too
seriously.  Yet, from the comparison with the ``$D$ sum rule'' we
conclude that the binding of $H$ is not as large as that expected in
the original quark models (without the instanton effects).
It should also be noted that the quark model generally
predicts the largest binding energy in the $SU(3)$ limit, \ie,
the symmetry breaking reduces the binding energy.
The sum rule predicts the opposite tendency.

\section{Conclusion}

The conclusion here is very short.  The theory predictions of the $H$
dibaryon mass have climbed from a deeply-bound ``6-quark exotic object''
to an unbound ``two baryons'',
while the experimental lower limit has increased.
In fact,  no definite prediction is yet given.
In a sense this is frustrating, but it can also be interpreted that
the $H$ dibaryon physics contains an essential part of QCD.
The hadron physics so far made a lot of predictions based on the
symmetry.  The best example would be the soft pion theorems for pion
dynamics.
They are quite robust because the chiral symmetry protects them.
Situation seems entirely different in baryon physics.
Most predictions there are model dependent, while few systematic
expansion methods are successful.
The fact that the $H$ predictions have big disparity among the models
indicates that
it contains an interesting physics.

Experimental searches are still strongly encouraged.

%
%
%
%
%
%

\end{document}